\documentclass[amsmath,amssymb,aps,twocolumn]{revtex4-1}

\usepackage{amssymb}
\usepackage{epsfig,graphicx}
\usepackage{graphicx}
\usepackage{amsmath}
\usepackage{color}
\usepackage{hyperref}
\hypersetup{
    colorlinks=true,
    linkcolor=blue,
    filecolor=magenta,      
    urlcolor=cyan,
    }

\newcommand{\be}{\begin{equation}}
\newcommand{\ee}{\end{equation}}

\newcommand{\bee}{\begin{eqnarray}}
\newcommand{\eee}{\end{eqnarray}}

\def\be{\begin{eqnarray} &&}

\def\ee{\end{eqnarray}}
\def\bew{\begin{widetext}}
\def\ew{\end{widetext}}

\usepackage{epsfig}

\def\bea{\begin{eqnarray}}
\def\eea{\end{eqnarray}}

\def\p {\pi}

\def\CP{CP }
\def\CPT{CPT }

\newcommand{\bppp}{$B^\pm \to \pi^\pm \pi^+\pi^- \,$}

\begin{document}
\title{ Enhanced Charm CP Asymmetries from Final State Interactions }
\author{I. Bediaga} 
\affiliation{Centro Brasileiro de Pesquisas F\'isicas, 
22.290-180 Rio de Janeiro, RJ, Brazil}
\author{T. Frederico}
\affiliation{Instituto Tecnol\'ogico de
Aeron\'autica, 12.228-900 S\~ao Jos\'e dos Campos, SP,
Brazil.}
\author{P. C. Magalh\~aes}
\email{p.magalhaes@cern.ch}
\affiliation{Instituto Tecnol\'ogico de
Aeron\'autica, 12.228-900 S\~ao Jos\'e dos Campos, SP,
Brazil.}
\affiliation{Departamento de Física Teórica and IPARCOS, Universidad Complutense de Madrid, E-28040, Madrid, Spain}
\date{\today}

\begin{abstract}
  We show that final state interactions (FSI) within a CPT invariant two-channel framework can enhance the charge-parity (CP) violation difference between  $D^0\to\pi^-\pi^+$ and $D^0\to K^-K^+$ decays up to the current experimental value.
This  result relies  upon: 
(i) the dominant tree level diagram, 
 (ii) the well-known experimental values for the $D^ 0\to\pi^-\pi^+$ and $D^ 0\to K^-K^+$ branching ratios,  and
 (iii)  the $\pi\pi \to \pi\pi$ and $\pi\pi \to K K $ scattering data to extract the  strong phase difference and inelasticity.   Based on well-grounded theoretical properties, we find the sign and bulk value of the $\Delta A_{CP}$ and  $A_{CP}(D^0 \to \pi^-\pi^+)$ recently observed by the LHCb Collaboration. 

\end{abstract}
\keywords{heavy meson, D meson three-body decay, CP violation,  final state interactions}
\maketitle

{\it Introduction.} 
Physics beyond the standard model (BSM), in general, predicts new sources of charge-parity violation (CPV). 
Experimentally, the high sensitivity and the clear signatures to observe the CPV in heavy meson decays
~\cite{BigiIJMP2020,Carla_ig,reviewLenz,Bigi:2021hxw,BigiSanda} led the search for these asymmetries to become an important branch of flavour physics. In particular, CPV in charm meson decay,  suppressed by the standard model, became a special tool 
to search for BSM effects, as suggested by Bigi and Sanda years ago, calling it ``The dark horse candidate"~\cite{BigiSanda}.

Recently the LHCb Collaboration made a significant step ahead in the understanding of CPV in charm, with the observation of the difference between the CP asymmetries of the singly Cabibbo-suppressed (SCS) $D^0 \to \pi^+ \pi^-$ and  $D^0 \to K^+ K^-$ decays~\cite{CPVCharm}:
\begin{eqnarray}
\Delta A^\text{LHCb}_{CP} &=&   A_{CP} (D^0 \to K^-K^+ )-A_{CP} (D^0 \to \pi^-\pi^+) \nonumber \\
&=& -(1.54 \pm 0.29)\times 10^{-3}.
\label{deltaacpexp}
\end{eqnarray}
This result is dominated by the direct CP asymmetry, with a negligible contribution from the $D^0 - \bar D^0$ oscillation~\cite{LHCbPRD2021}. The observed value of $\Delta A_{CP}$ was understood to be at the borderline of the Standard Model and  BSM  interpretations~\cite{reviewLenz}. %
The world average is ~\cite{HFLAG_D0pp}:
\begin{equation}\label{deltaacpave}
\Delta A^\text{av}_{CP} = - (1.61 \pm 0.28) \times 10 ^{-3}\, ,
\end{equation}
with the channel asymmetries  defined as:
\begin{equation}\label{eq:Acp}
 A_{CP}(f)=\frac{\Gamma\left(D^0\to f\right)-\Gamma(\bar D^0\to f)}{ \Gamma\left(D^0\to f\right)+\Gamma(\bar D^0\to f) 
}\, ,
\end{equation}
where $f$ represents the final state.

There are several theoretical frameworks that address CPV in charm within the Standard Model. They can be divided between those using the QCD short-distance approach~\cite{Petrov2017,Chala2019} and those considering long-distance effects through FSI~\cite{Grossman2019,Soni2022}, including the topological approach with $SU(3)$ breaking~\cite{Silvestrini, Uli_Schacht,BucellaPRD2019,ChengPRD2019}. In the charm sector, QCD has known problems to access the charm penguin contribution and the QCD based approach~\cite{Petrov2017} has predicted $\Delta A_{CP}$ one order of magnitude lower than the experimental value. On the other hand, the available long-distance approach tries to explain the CPV result in charm  by exploring model-dependent fitting to nonperturbative aspects of QCD.  
\begin{figure}[!htb] 
\vspace*{-5mm}\includegraphics[width=4.5cm]{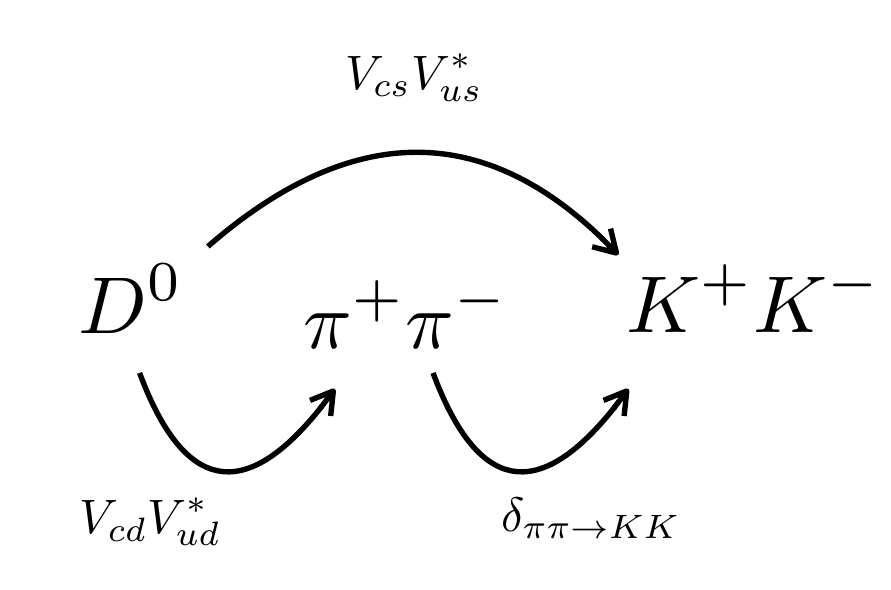} 
\vspace*{-3mm}
\caption{ Illustration of the mechanism for direct CPV in $D^0$ (and $\bar{D^0}$) decays  driven by  $\pi^+\pi^- \to K^+K^-$ rescattering.}  \label{D0ppkk}\end{figure}

The importance of FSI in charm decays has been known for a long time~\cite{Biswas, Buccella_Lusignoli,PRD84} but only recently used to investigate CPV effects \cite{Soni2022, Grossman2019}. 
In this work, we go beyond the previous FSI analysis. Within a CPT conserving framework, where the total width of the particle and antiparticle should be the same \cite{wolfenstein},  and considering the rescattering process $\pi^+\pi^- \to K^+K^-$, one can produce the interference necessary to magnify the CPV in the $D^0 \to \pi^-\pi^+$ and  $D^0 \to K^-K^+ $ amplitude decays. For the meson scattering amplitudes, we  use the values observed in the 1980s~\cite{Cohen,Brookhaven1}.
Such a mechanism is illustrated in Fig.~\ref{D0ppkk} and can explain the sign and bulk values of $\Delta A_{CP}$ and the $A_{CP}(D^0 \to \pi^-\pi^+)$ observed recently by LHCb  \cite{CPVCharm, LHCbcharmNEW}.

The interference mechanism between $\pi^+\pi^-$ and $K^+K^-$ states due to the strong final state interaction (FSI) in the S wave, was also shown to explain the large amount of CPV observed in some regions of the phase space of charmless three-body $B$ decays~\cite{BediagaPRD2014,Nogueira2015,TIPD_Novo}, as reviewed in~\cite{Carla_ig}. In $D$ decays, this idea is also present in  Grossman and Schacht~\cite{Grossman2019} within symmetry approach.

\begin{figure}[!thb]
  \includegraphics[width=8.cm]{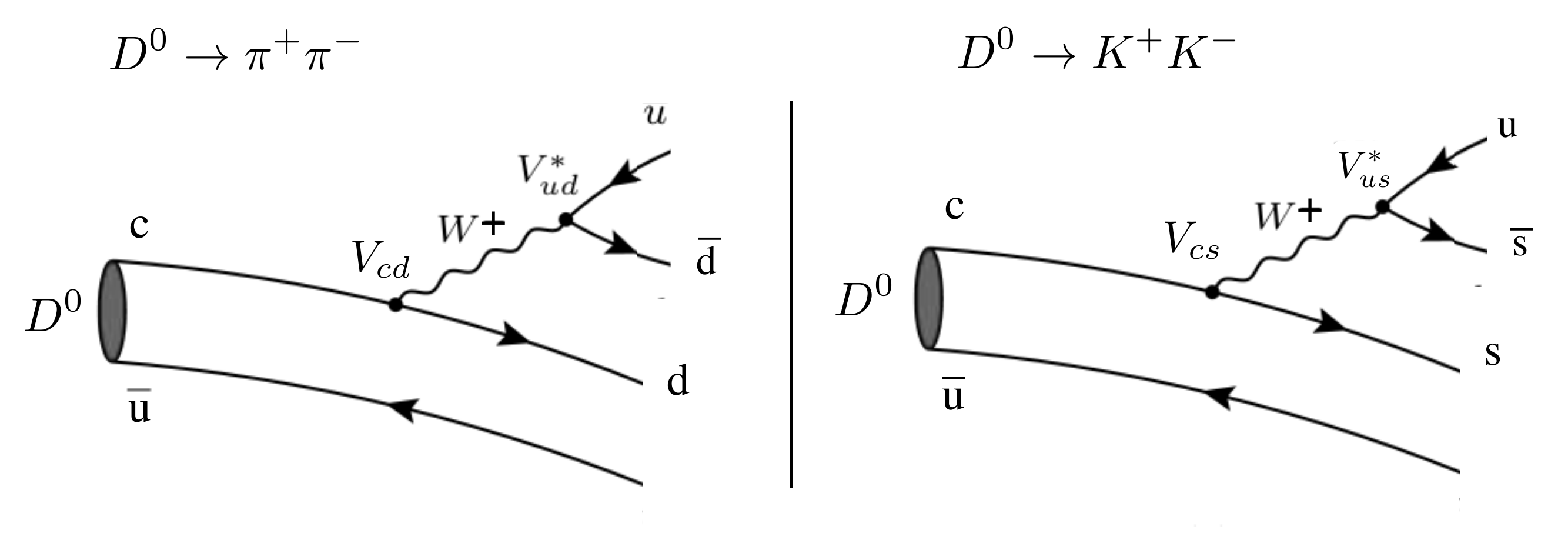}
  \vspace*{-2mm}
  \caption{ Quark tree diagrams for the $D^0\to\pi^+\pi^-$ and $D^0 \to K^+K^-$ decays.}
  \label{D0pp}
\end{figure}

Here we only consider contributions  from tree level diagrams
, as given in Fig.~\ref{D0pp}, and build the corresponding decay amplitudes with well-grounded properties of the SM: 
(i) the CPT invariance assumption relating decays with the same quantum numbers; 
(ii) the Watson theorem relating the  strong phase of the rescattering process $\pi^+\pi^- \to K^+K^-$ to the decay amplitudes; and 
(iii) the unitarity of the strong S matrix. 

{\it CPT implications for CPV.} The CPT constraint has been used in charmless  B decays with exciting results for experimental analysis~\cite{LHCbPRL2020,LHCbPRD2020} and phenomenological interpretations~\cite{BediagaPRD2014,Nogueira2015,TIPD_Novo, PatBnoCPV}. The large phase space available in B decays allows, in principle, several possible rescattering contributions for each channel, which brings into question the CPT invariance constraint. 
However, this argument does not hold for charm meson decays with a small and well-explored phase space.

The final states of nonleptonic SCS $D^0$ decays involves only mesons, with the dominance of pion and kaon  $(M)$ channels. In principle, the FSI could mix all these states, through the general strong S matrix, involving any number of mesons, allowed by the phase space: \begin{equation}{\small \label{eq:stotal}
S=\begin{pmatrix}
S_{2M,2M} &  S_{2M,3M} &S_{2M,4M}&\cdots\\
 S_{3M,2M} & S_{3M,3M} &S_{3M,4M}& \cdots
 \\
 S_{4M,2M} & S_{4M,3M} &S_{4M,4M}& \cdots
  \\
 \cdots &  \cdots & \cdots & \cdots
\end{pmatrix}}
\, ,
\vspace*{-1mm}
\end{equation}
where each element is a matrix representing the strong coupling between the channels with a number of mesons $n$, labeled by $(nM)$.
In particular, considering the final state interactions in $D^0 \to \pi^+\pi^-$ and  $D^0 \to K^+K^- $ decays,  we know that two pions cannot 
 branch to three pions due to the G parity.
For the purpose of finding the main mechanism that drives CPV,
 we can ignore four pion coupling to the $2M$ channel, namely $S_{2M,4M}\approx 0$ and $S_{4M,2M}\approx 0$ (based on $1/N_c$ counting arguments~\cite{Buras:1985xv,Smith:2003iu}), and the coupling to the $\eta \eta$ channel~\cite{Lindenbaum:1991tq}, considering that their couplings to the $\pi\pi$ channel are suppressed with respect to the $K\bar{K}$ one. 

 We remark that although the $D^0\to 4\pi$ decays have a large branching fraction, there is no clear evidence of strong coupling between the two and four pion channels in the $D^0$ mass region.  The only observable that decays in both channels is the $f_0(1500)$ scalar resonance. For other scalar states such as $f_0(980)$ and $f_0(1710)$, the dominant channels are $K\bar K$ and two pions, with no observation of four pions reported. 

Consequently, for CPV studies of $D^0 \to \pi^+\pi^-$ and  $D^0 \to K^+K^- $ decays,  it is a good approximation to consider $S_{2M,2M}$ restricted to the $(\pi\pi,KK)$ S-wave channels. 
 The $S_{2M,2M}$ unitarity leads to an important CPT constraint in which
 the $A_{CP}$'s must have opposite signs.

{\it FSI and CPT constraint.} 
If we assume that the single Cabibbo suppressed $D^0 \to \pi^- \pi^+$ and  $D^0 \to K^- K^+$  decays proceed via tree level amplitudes, neglecting the suppressed contribution from penguins ($P/T\sim 0.1$ \cite{reviewLenz}), as depicted in Fig.~\ref{D0pp}, there is no possibility to generate CP violation other than coupling these two channels, which have different weak phases, via the strong interaction. 
This is fulfilled by the rescattering mechanism explicitly illustrated in Fig.~\ref{D0ppkk}. 

The weak phase difference comes from the CKM matrix elements in the tree amplitudes of Fig.~\ref{D0pp},
with the  CP violating phase carried by
$V_{cd}V^*_{ud}$.  The weak phase in $V_{cs}V^*_{us}$ was neglected, as it is much smaller
than the one in $V_{cd}V^*_{ud}$~\cite{PDG}.

\begin{figure}[!htb]
\vspace{-3mm}
 \includegraphics[width=3.5cm]{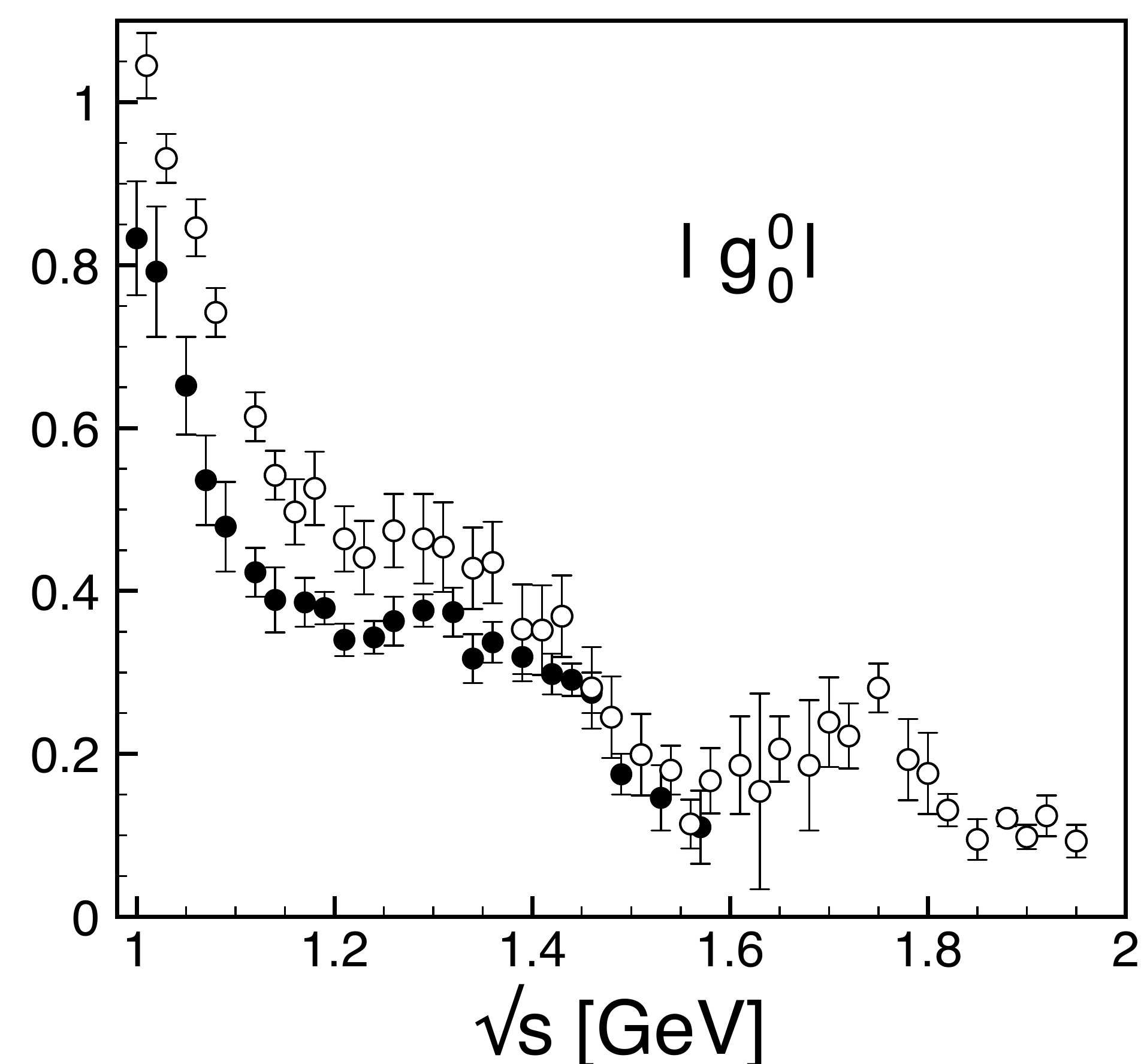}
 \includegraphics[width=3.5cm]{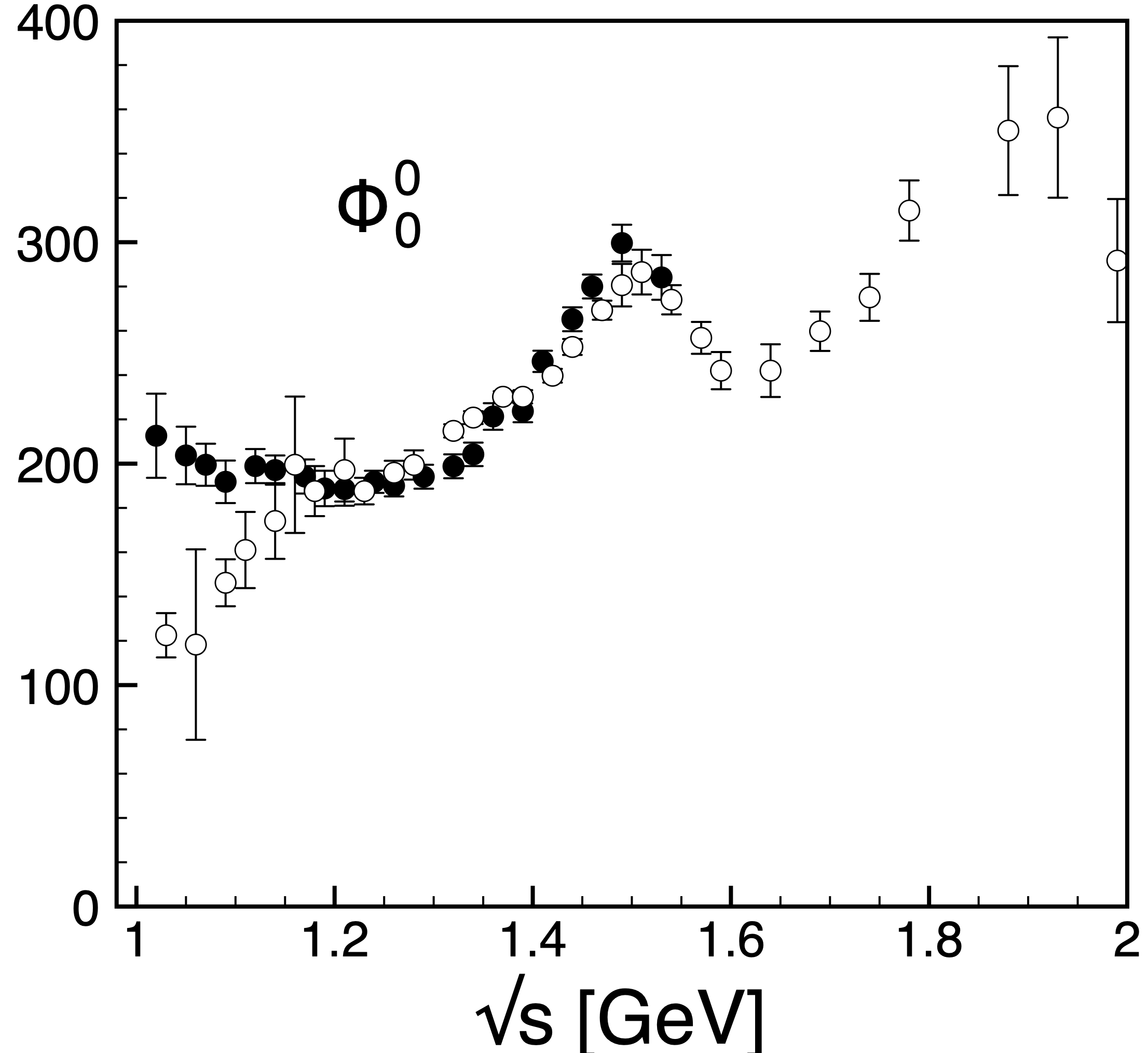}
 \vspace*{-1mm}
  \caption{$\pi\pi\to K\bar{K}$ amplitude $|g^0_0|$ (left) and phase $\Phi^0_0$ in degrees  (right), associated with  $S_{\pi\pi,KK}$ Eq.~\eqref{eq:s2p2k}.   Experimental data from Argonne~\cite{Cohen} (full) and Brookhaven~\cite{Brookhaven1} (empty).}
  \label{Pelaez}
  \vspace{-3mm}
\end{figure}

The Watson theorem says that the strong phase $\delta_{\pi\pi \to KK}$ is the same, independent of the initial process. Therefore, we can use the parameters obtained in the $\pi\pi$ scattering from the $\pi N \to \pi \pi N$ and $\pi N \to KK N$ reactions~\cite{CERN-Munich75,Cohen, Brookhaven1}. These experiments observed two important properties in the $\pi^+\pi^- $ S wave:  the inelasticity parameter decreases drastically  above 1~GeV with the opening of the KK channel~\cite{CERN-Munich75};
and the $K K$ channel dominates the inelasticity of the $\pi^+\pi^- $ one below 2~GeV, which also supports our previous discussion. To be more concrete,
in Fig.~\ref{Pelaez} we show two experimental results for the $\pi\pi \to K\bar K$
scattering amplitude  in the scalar-isoscalar  state.
From this figure, we can extract the transition amplitude needed to compute the rescattering effects in the $D^0 \to \pi^+\pi^-$ and  $D^0 \to K^+K^- $  decays  near the $D_0$ mass.

For our purpose, it is enough to know  the S-wave S matrix  for  the coupled channels
  $\pi^-\pi^+$ and $K^-K^+$:
\begin{equation}\label{eq:s00}
{\small
S_{2M,2M}=\begin{pmatrix}
S_{\pi\pi,\pi\pi} &  S_{\pi\pi,K K} \\
 S_{KK,\pi\pi} & S_{K K,K K}
\end{pmatrix}}
\, ,
\end{equation}
where $S_{\pi\pi,\pi\pi}=\eta\, \text{e}^{2i\delta_{\pi\pi}}$, $S_{KK,KK}=\eta\, \text{e}^{2i\delta_{KK}}$ and
$S_{\pi\pi,K K}=S_{KK,\pi\pi}=\imath \sqrt{1-\eta^2}\,\text{e}^{\imath (\delta_{\pi\pi}+\delta_{KK})}$, with
$\delta_{\pi\pi}$ and $\delta_{KK}$ the elastic phase shifts, and  $0\le\eta\le 1$ the absorption parameter.
To quantify $\eta$, we used 
the parametrization of the off-diagonal
S-matrix element from~\cite{Pelaez-Rodas,Pelaez19}:
\begin{equation}\label{eq:s2p2k}
    S_{\pi\pi,K K}(s)
    =i\,4\,\sqrt{\frac{q_\pi q_K}{s}}|g^0_0(s)|\,e^{i\phi^0_0(s)}\, {\Theta(s-4m^2_K)}
    \, ,
\end{equation}
where $\phi^0_0=\delta_{\pi\pi}+\delta_{KK}$,  $q_\pi=\frac12\sqrt{s-4m_\pi^ 2}$ and  $q_K=\frac12\sqrt{s-4m_K^ 2}$.
From  Fig.~\ref{Pelaez} one finds that, at the $D^0$ mass $|g^0_0(M^2_D)|= 0.125\pm0.025$, which from
~\cite{Pelaez-Rodas,Pelaez19} gives $\sqrt{1-\eta^2}= 0.229\pm0.046$ and $\eta= 0.973\pm0.011$. Also, we have $\phi^0_0(M^2_D)=
343^\text{o}\pm8^\text{o}$.

 Summarizing our assumptions up to this point, we: i) ignored the sub-leading diagrams of the amplitude decay; ii) considered the dominant FSI in $\pi\pi$ to be the $ KK$ channel; and iii) used a data driven approach to extract both $\pi\pi$ and  $ \pi\pi \to KK$ magnitude and phases at the $D_0$ mass energy. With these assumptions,
the total $D^0$ decay amplitudes produced by the tree diagrams of Fig.~\ref{D0pp} are dressed by the hadronic FSI and receive contribution from both diagonal and off-diagonal S-matrix elements from Eq.~(\ref{eq:s00}). The resulting amplitude is denoted by $\mathcal{A}_{D^0\to f}$, with $f$ labeling the $0^+$ final states restricted to the $f\equiv$ $\pi^+\pi^-$ and $K^ +K^-$ channels: 
\begin{equation}
    \begin{aligned}
\mathcal{A}_{D^0\to KK}=&\eta\, \text{e}^{2i\delta_{KK}}\,V^*_{cs}V_{us}\,a_{KK}  \\ & +\, i\sqrt{1-\eta^2}\,\text{e}^{i(\delta_{\pi\pi}+\delta_{KK})}\,V^*_{cd}V_{ud}\,a_{\pi\pi}\, , \\
\mathcal{A}_{D^0\to\pi\pi}=& \eta \,\text{e}^{2i\delta_{\pi\pi}}\,
V^*_{cd}V_{ud}\,a_{\pi\pi}  \\ & + \,i\sqrt{1-\eta^2}\,\text{e}^{i(\delta_{\pi\pi}+\delta_{KK})}\,V^*_{cs}V_{us}\,a_{KK}\,.
\end{aligned}
\label{eq:cp1}
\end{equation}
For the $\bar D^0\to f$  decay amplitude, $\mathcal{A}_{\bar D^0\to f}$, the CKM matrix elements are their complex conjugates.  

The amplitudes $a_{KK}$ and $a_{\pi\pi}$ do not carry any strong or weak phases, due to the tree level nature of the decay process. All of the hadronic FSI comes from S-matrix elements that has been factored out and included in the  $D^0$ and $\bar D^0$ decay amplitudes.  Equation.~\eqref{eq:cp1}  is equivalent to the leading order amplitudes due to the strong interaction derived in~\cite{BediagaPRD2014} and
based on Refs.~\cite{wolfenstein,BigiSanda}. 

The CPT constraint restricted to the two-channels corresponds to:
\begin{equation}
\sum_{f=(\pi\pi,KK)}(|\mathcal{A}_{D^0\to f}|^2-
|{\mathcal{A}}_{\bar D^0\to f}|^2)=0\, ,
\label{eq:deltacpt8}
\end{equation}
which is fulfilled by the proposed decay amplitudes of Eq.~\eqref{eq:cp1} and their charge conjugate ones. It is worth noting that the essential ingredients to derive the result shown in \eqref{eq:deltacpt8} are the unitarity of the S matrix of the two-channel model and the weak phase assigned by the products of the CKM matrix elements.
 One could write the analogous of Eq.~\eqref{eq:deltacpt8} including more  strongly coupled channels. However, as we argued before, we want to investigate the main mechanism and so we keep only the dominant $(\pi\pi,KK)$ channels.  

The identity expressed by \eqref{eq:deltacpt8} illustrates how the so called
``compound" \CP asymmetry~\cite{Atwood1998,Soni2005}, including the effects of the weak and strong phases, has the important property of the two terms canceling one other, when summed over all final states, in order to satisfy the \CPT condition.

{\it CP asymmetries in $D^0\to \pi^-\pi^+$ and $D^0\to K^-K^+$.} 
 The CPV difference in the partial decay widths of $D^0$ and $\bar D^0$
 is defined as 
$ \Delta \Gamma_f=\Gamma\left(D^0\to f\right)-\Gamma(\bar D^0\to f)\,$. 
 By considering the amplitudes in Eq.~\eqref{eq:cp1} and those for the charge conjugate state, we get the following:
\begin{equation}
\begin{aligned}
\Delta \Gamma_{\pi\pi}=-\Delta \Gamma_{KK}
&=4\,\mbox{Im}[V_{cs}V^*_{us}V^*_{cd}V_{ud}]
 \\
&\times a_{\pi\pi}\,a_{KK}\,
\eta\sqrt{1-\eta^ 2} \cos\phi
\, ,
\end{aligned}
\label{cp26-3}
\end{equation}
where $\phi=\delta_{KK}-\delta_{\pi\pi}$,
remembering that  $a_{\pi\pi}$ and $a_{KK}$ are real and have the same sign.  
Note that the sign of $\Delta \Gamma_{f}$ is determined by the elements of the CKM matrix and the elastic S-wave phase-shifts in the two final state channels.

In order to obtain the $A_{CP}$'s, one has to estimate  $a_{\pi\pi}$ and $a_{KK}$, which can be done using the partial widths of the decays $D^0\to \pi^+\pi^-$ and $D^0\to K^+K^-$, extracted from
 the amplitudes given in Eq.~\eqref{eq:cp1}.
 Assuming 
that 
$\sqrt{1-\eta^ 2}\ll 1$ at the $D^0$ mass, we have:
\begin{small}
\begin{equation}\label{eq:Gamma}
\Gamma_{\pi\pi}\approx
\eta^2 |V^*_{cd}V_{ud}|^2\,
a_{\pi\pi}^2 
~\,\,\text{and}~\,\,
\Gamma_{KK}\approx \eta^2
|V^*_{cs}V_{us}|^2 a _{KK}^2\, .
\end{equation}
\end{small}
If we use the branching fraction information $\text{Br}(D^0\to \pi\pi)$ and $\text{Br}(D^0\to KK)$, we can determine $a_{\pi\pi}$ and $a_{KK}$ with Eq.~\eqref{eq:Gamma}.  

Here,
we do not address the values of the branching ratios and assume they
are consistent with the standard model.
Our data 
driven approach
tackles only the difference between the two decay widths and we use
the experimentally measured values of the branching ratios as inputs
for the computation of $A_{CP}$.
 We observe that the branching ratio values have been the subject of investigation in the literature\cite{ChengPRD2019,Buccella_Lusignoli,Biswas}.

The  CP asymmetries are then, from  Eqs.~\eqref{cp26-3} and~\eqref{eq:Acp},  given by:
\begin{equation}
\begin{aligned}
&A_{CP}(f)
\approx  \pm 2\,{\mbox{Im}[V_{cs}V^*_{us}V^*_{cd}V_{ud}]
\over|V_{cs}V^*_{us}V^*_{cd}V_{ud}|} \,
\\
&\times \eta^{-1}\,\sqrt{1-\eta^ 2}\, \cos\phi\, \Bigg[{ 
\,\,\,\text{Br}(D^0\to K^+K^-)
\over 
\text{Br}(D^0\to\pi^+\p^-)}
\Bigg]^{\pm\frac12}
, 
\end{aligned}
\label{cp26-4}
\end{equation}
where $+$ and $-$ stand for $f=\, \p^+\p^-$ and $K^+K^-$, respectively,
and the CKM factors 
ratio reads~\cite{PDG}
\begin{eqnarray}
{\mbox{Im}[V_{cs}V^*_{us}V^*_{cd}V_{ud}]
\over|V_{cs}V^*_{us}V^*_{cd}V_{ud}|}=(6.02\pm 0.32)\times10^{-4}
\, .
\label{eq:deltaCPf}
\end{eqnarray}

{\it Estimation of the CP asymmetries.}
Inspecting the CP asymmetry in Eq.~\eqref{cp26-4}, the remaining unknown quantity is the difference between  the $KK$ and $\pi\pi$ S-wave phase shifts.   Ideally, to  quantify the contribution from $\cos(\delta_{KK}-\delta_{\pi\pi} )$  at the $M_D$ energy, we could  directly inspect the phase data at this point.
However, differently from $\pi\pi$, there is no $K\bar{K}$ scattering data from meson-nucleon interactions. 
Without precise knowledge of the $K\bar{K}$ phase, we  use $\delta_{KK}-\delta_{\pi\pi}\,=\,(\delta_{KK}+\delta_{\pi\pi})-2\delta_{\pi\pi}\,=\,\phi^0_0-2\delta_{\pi\pi}$. 
From $\pi\pi$ scattering data~\cite{CERN-Munich75,Ochs2013}  and  the $\pi\pi \to KK$ phase, given in Fig.~\ref{Pelaez}, we obtained
$\cos(\delta_{KK} -\delta_{\pi\pi} )\lesssim 1$ in the high mass region. This can be verified  from $1.58$ to $1.78$~GeV, the  upper limit of the data~\cite{CERN-Munich75}. The  $\cos \phi$'s extracted from the updated CERN-Munich data for $\delta_{\pi\pi}$ ~\cite{Ochs2013} and 
$\phi^0_0=\delta_{KK}+\delta_{\pi\pi}$ from~\cite{Pelaez-Rodas} are very close to 1. At  $M_{D^0}$ energy, $\delta_{\pi\pi}$ comes from the extrapolation given in~\cite{Pelaez19} (Solution II, which is consistent with the data~\cite{CERN-Munich75} and \cite{Ochs2013}), resulting in $\cos\phi= 0.99 \pm  0.18  $.
Note that at this energy the parametrization ~\cite{Pelaez19} has a large error bar. 

Given 
the branching fraction values~\cite{PDG}:
\begin{equation}
\begin{aligned}
&\text{Br} (D^0\to \pi^+\pi^- ) 
=(1.455\pm0.024)\times 10^{-3}\, ,
\\
&\text{Br} (D^0\to K^+K^-) = ~ (4.08\pm0.06)\times 10^{-3}\, .
\end{aligned}
\label{eq:GPPGKK}
\end{equation}
 all parameters for calculating the CP asymmetries, of Eq.~\eqref{cp26-4},  are well defined, except for $\eta$. So we factorize its dependence as:
\begin{equation}
    \begin{aligned}
&A_{CP}(\pi\pi)  \,\,\,\,=\,\,\, \, \,  (1.99\pm 0.37)\times 10^{-3}\,\sqrt{\eta^{-2}-1}\, ,\\
&A_{CP}(KK)  \,=\,   -(0.71\pm0.13)\times 10^{-3}\,\sqrt{\eta^{-2}-1}\,,
\label{Acpth}
    \end{aligned}
\end{equation}
and from that:
\begin{equation}
    \Delta A^{th}_{CP}=-(2.70\pm 0.50)\times 10^{-3}\,\sqrt{\eta^{-2}-1}\, .
\end{equation}

  As we pointed out earlier in Fig.~\ref{Pelaez}, there is only one datum for $\pi\pi \to KK$ with center mass energy above 1.8~GeV, needed to reach the $D^0$ mass. The solution gives $\eta\approx 0.973\pm0.011$~\cite{Pelaez-Rodas},  which implies 
 \bea
 \Delta A^{th}_{CP} =
  -(0.64\pm 0.18) \times 10^{-3}\, .
  \label{DeltaAcp}
  \eea 
 The result we found for $\Delta A^{th}_{CP} $ clearly shows the relevant enhancement of FSI for this quantity, arriving at the sign and bulk value of the LHCb observation. This indeed is the largest theoretical prediction within SM without relying on fitting parameters~\cite{reviewLenz}.
 
 Although the systematic uncertainties are absent in $|g^0_0|$ seen in Fig.~\ref{Pelaez}, the experimental study used to extract these values at high energies~\cite{Brookhaven1}, reported high systematic uncertainties in their estimate of other experimental parameters obtained in that analysis.  Therefore the quoted error in $\eta$ in this case is underestimated, which impacts the error in Eq.~\eqref{DeltaAcp}.
 
In order to explore other possible values of the inelasticity, if instead of using the $\pi\pi \to KK$ data, one uses $\pi\pi \to \pi\pi$ from Grayer et al.~\cite{CERN-Munich74}, one finds $\eta =0.78\pm 0.08$. 
In this case, the
approximation $\sqrt{1-\eta^2} \ll 1$ does not hold and we considered the complete solution of Eq.~\eqref{eq:cp1}. In order to keep $a_{\pi\pi}$ and $a_{KK}$ real, we choose  $\delta_{\pi\pi}-\delta_{kk}=30^\text{o}$, that is within the quoted error for $\cos \phi$.
That gives
\begin{equation}\label{eq:deltath2}
\Delta A^{th}_{CP} =  
(-1.31\pm 0.20)\times 10^{-3} \,. 
  \end{equation}
  This value is compatible with the LHCb experimental results within $1\sigma$, and relies on our assumption that the $K\bar{K}$  channel saturates the inelasticity in $\pi\pi$ scattering at the $D^0$ mass.
  
   Independently of the 
  value for $\eta $, 
   we can make a prediction for future experimental results of the ratio:
 \begin{small}
$$\frac{A_{CP}(D^0\to \pi^-\pi^+)}{A_{CP}(D^0\to K^-K^+)}=-\frac{\text{Br}(D^0\to K^-K^+)}{\text{Br}(D^0\to \pi^-\pi^+)}=-2.8\pm0.06\, .
$$ 
\end{small}

In fact, relying only on the CPT constraint for two channels, given by Eq.~\eqref{eq:deltacpt8}, 
one can easily obtain the CP asymmetries as follows:
\begin{small}
\begin{equation}
\label{eq:cptpredict1}
    \begin{aligned}
    &A_{CP}(\pi\pi)=-\frac{\Delta A_{CP}\,\text{Br} (D^0\to K^+K^-)}{\text{Br} (D^0\to K^+K^-)+\text{Br} (D^0\to \pi^+\pi^- ) }\, , \\
     &A_{CP}(KK)=\frac{\Delta A_{CP}\,\text{Br} (D^0\to \pi^+\pi^- ) }{\text{Br} (D^0\to K^+K^-)+\text{Br} (D^0\to \pi^+\pi^- ) } \, ,
    \end{aligned}
\end{equation}
\end{small}
\noindent which are also valid for the $A_{CP}$'s from Eq.~\eqref{cp26-4}. Using  experimental inputs for $\Delta A_{CP}$ and Br's  we predict the values for the $A_{CP}$'s:
\begin{equation}\label{eq:cptpredict}
\begin{aligned}
&A_{CP}(\pi\pi)~~
=~~(1.135\pm 0.021)\times10^{-3}\, ,
\\
&A_{CP}(KK)=-(0.405\pm 0.077)\times10^{-3}
\, .
\end{aligned}
\end{equation}

{\it Summary.} 
We predict an enhancement of the $A_{CP}$'s and $\Delta A_{CP}$ for the SCS decays $D^0(\bar D^0)\to \pi^-\pi^+$ and $ D^0(\bar D^0)\to K^-K^+$, relying solely on SM physics. 
The enhancement is a consequence of $\pi^+\pi^-$ and $K^+K^-$ coupling via the FSI,  whose strong phase contribute to both amplitudes with opposite sign,  due to CPT invariance. 
Our approach  takes into account the final state interaction in accordance with the Watson theorem, besides the standard CKM matrix elements. 
If our prediction for the $A_{CP}$'s ratio is  confirmed, the forthcoming data could constrain the S-wave phase-shift difference  in the $\pi^+\pi^-$ and $K^+K^-$ elastic  channels at the $D^0$ mass, as well as  the magnitude of the off-diagonal S matrix.

Very recently, during the revision process of this work, the LHCb Collaboration presented new results for $D^0(\bar D^0)\to \pi^-\pi^+$ and $ D^0(\bar D^0)\to K^-K^+$ \cite{LHCbcharmNEW} which confirms our prediction that 
$|A_{CP} (\pi\pi)|> |A_{CP} (KK)|$:
\begin{equation}
\begin{aligned}
\label{eq:LHCbnewCP}
&A^{LHCb}_{CP} (\pi\pi) =(2.32 \pm 0.61)\times 10^ {-3}\, , \\  &A^{LHCb}_{CP} (KK) =(0.77 \pm 0.57)\times 10^ {-3}\, ,
  \end{aligned}
  \end{equation}
 with the result for $\pi\pi$ channel being the first evidence of an individual charm decay asymmetry. Note that both LHCb new $A_{CP}$  values are statistically compatible with  ours results.  From Eq.s~\eqref{eq:deltath2}   and \eqref{eq:cptpredict1},
 we find the central values $A_{CP} (\pi\pi) =(0.97  \pm 0.05)\times 10^ {-3} $ and
$A_{CP} (KK) =-(0.34  \pm 0.15)\times 10^ {-3}$ .
These values are compatible with the experimental ones within 2$\sigma$s and also  with the results given by Eq.~\eqref{eq:cptpredict}, obtained from  $\Delta A^{LHCb}_{CP}$ and the experimental branching ratios.  
 If the hint
for a positive $A_{CP}(KK)$ is confirmed by a more precise measurement,
the scenario presented here would be disfavored as a solution to the
 $\Delta A_{CP}$ puzzle.

 The same rescattering mechanism can contribute to CPV in three-body SCS $D$ decays.
In fact, one expects that the CP asymmetry must be enhanced in the three-body  $D^+ \to \pi^+ \pi^-\pi^+$ and  $D^+ \to K^+ K^- \pi^+$ phase-space distribution~\cite{BigiReis2014}, where the $\pi^+\pi^- \to K^+K^-$ rescattering is relevant in a large fraction of the phase space available to $K^+K^-$, as seen in Fig.~\ref{Pelaez}. This is left for future study. 

 Furthermore, as pointed out several times~\cite{BigiIJMP2020,Bigi:2021hxw}, the SM gives almost no contribution to CPV in double Cabibbo suppressed (DCS) decays. If CPV is observed in DCS modes, this will  point to new physics.  Following the present approach, the best channels to observe CPV in DCS, are the $D^+ \to K^+\pi^-\pi^+$ and  $D^+ \to K^+K^-K^+$, which also has the rescattering $\pi^-\pi^+ \to K^-K^+$ as a mechanism to enhance the observable CP violation.  

{\it Acknowledgments.} 
We would like to thank J. R. Pelaez for clarifying discussion and, along with A. Rodas, providing results from their parametrization. We also thank T. Pajero and  B. V. Carlson for helping improve the manuscript. This study was financed partly by Conselho Nacional de Desenvolvimento Cient\'{i}fico e Tecnol\'{o}gico (CNPq) under the Grant No. 308486/2015-3 (TF) and INCT-FNA Project No. 464898/2014-5,   FAPESP Thematic Projects Grants No. 2017/05660-0 and No. 2019/07767-1 (TF), FAPERJ, CAPES,  CAPES - PRINT Grant No. 88887.580984/2020-00 and Spanish Ministerio de Ciencia e Innovaci\'on Grant Maria Zambrano para atracción de talento interancional (Convocatoria 2021-2023). P.C.M. would like to thank University of Bristol for the support as well.
This research has benefited from the support of the Munich Institute for Astro-Particle and BioPhysics (MIAPbP), which is funded by the Deutsche Forschungsgemeinschaft (DFG, German Research Foundation) under Germany Excellence Strategy – EXC-2094 – 390783311.

\newpage


\end{document}